%% file: main.tex
\documentclass[conference]{IEEEtran}
\IEEEoverridecommandlockouts
\usepackage{cite}
\usepackage{amsmath,amssymb,amsfonts}
\usepackage{algorithmic}
\usepackage{graphicx}
\usepackage{textcomp}
\usepackage{xcolor}
\usepackage{multirow}
\usepackage{tabularx}
\usepackage[normalem]{ulem}
\useunder{\uline}{\ul}{}
\usepackage{float}

\input{alphabet}
\def\T{^\mathsf {T}}

\begin{document}
\bstctlcite{IEEEexample:BSTcontrol}  
\title{Compact Implicit Neural Representations for Plane Wave Images
}

\author{
\IEEEauthorblockN{
Mathilde Monvoisin\IEEEauthorrefmark{1}\IEEEauthorrefmark{2},
Yuxin Zhang\IEEEauthorrefmark{1}\IEEEauthorrefmark{2} 
and Diana Mateus\IEEEauthorrefmark{1}}
\thanks{\IEEEauthorrefmark{2}Equal contribution}
\IEEEauthorblockA{\IEEEauthorrefmark{1}Nantes Université, École Centrale Nantes, LS2N,
CNRS, UMR 6004, F-44000 Nantes, France \\
Email: mathilde.monvoisin@ls2n.fr, yuxin.zhang@ls2n.fr}
}


\maketitle

\begin{abstract}
\input{sections/0_abstract}
\end{abstract}


\section{Introduction}
\input{sections/1_introduction}

\section{Methods}
\label{sec: methods}
\input{sections/2_methods}

\section{Experiments}
\label{sec: experiments}
\input{sections/3_experiments}

\section{Conclusion}
\input{sections/4_conclusion}

\section*{Acknowledgment}
\input{sections/acknowledgment}

\bibliographystyle{IEEEtran} 
\bibliography{biblio}

\vspace{12pt}

\end{document}

%% file: sections/0_abstract.tex
Ultrafast Plane-Wave (PW) imaging often produces artifacts and shadows that vary with insonification angles. We propose a novel approach using Implicit Neural Representations (INRs) to compactly encode multi-planar sequences while preserving crucial orientation-dependent information.
To our knowledge, this is the first application of INRs for PW angular interpolation. Our method employs a Multi-Layer Perceptron (MLP)-based model with a concise physics-enhanced rendering technique. Quantitative evaluations using SSIM, PSNR, and standard ultrasound metrics, along with qualitative visual assessments, confirm the effectiveness of our approach.
Additionally, our method demonstrates significant storage efficiency, with model weights requiring 530 KB compared to 8 MB for directly storing the 75 PW images, achieving a notable compression ratio of approximately 15:1.


%% file: sections/1_introduction.tex
Recent studies have demonstrated that Implicit Neural Representations (INRs) are highly effective for continuously approximating both scalar and vector fields. In the field of Computer Vision, INRs have proven adept at accurately representing intricate 3D scenes and generating novel views~\cite{nerf, dbarf, wire, siren}.

There has been increasing interest in using INRs for ultrasound in the recent years. For instance,  with freehand 3D ultrasound datasets, Gu~\cite{gu2022INR} and Yeung~\cite{yeung2021INR} trained neural networks to model 3D volumes, mapping coordinates to grayscale voxel intensities. Additionally, Wysocki et al.~\cite{ultra-nerf} utilized a ray-tracing-based neural rendering technique to learn tissue properties from B-mode images. 
In the context of vascular modeling, Alblas et al.~\cite{alblas2022INR} employed INRs to fit multiple nested surfaces of abdominal aortic aneurysms. Velikova~\cite{velikova2023INRaorta} and Song~\cite{song2022INRcarotid} used semantic segmentation data to reconstruct the aorta and carotid vessels, respectively. Furthermore, Li et al.~\cite{li2021INRspine} applied the Neural Radiance Field (NeRF)~\cite{nerf} algorithm to reconstruct spinal structures. In this work, we focus on reconstructing multi-angle Plane-Wave (PW) ultrasound images. Specifically, a neural network is trained to map positional and angular information to pixel intensities.

Ultrafast imaging modalities, such as PW beamforming, have gained significant attention over the past decade~\cite{pw2009,pw2015}. However, each ultrasound image reconstructed from a single PW transmission using the conventional Delay-And-Sum (DAS) method~\cite{DAS} often exhibits artifacts or shadows. These artifacts vary in pattern depending on the angle of insonification.
The angle dependency of these artifacts is particularly intriguing because it can provide valuable insights into tissue characteristics, thereby enhancing diagnostic accuracy.
In this study, we propose adapting INRs in the context of Multi-Plane-Wave acquisitions. Our objective is to develop an INR capable of compactly encoding a multi-planar acquisition sequence while preserving orientation-dependent information, particularly in regions affected by shadows or artifacts. Once the neural network is well-trained on a sequence of tiled PW images, it can not only reconstruct images from the training set but also infer new views using arbitrary image grids and PW angles.

Despite numerous studies utilizing INRs for ultrasound and the work by Afrakhteh et al.~\cite{afrakhteh2021, afrakhteh2023} employing tensor completion for PW angular interpolation, to the best of our knowledge, this is the first study to use INRs to achieve PW angular interpolation.

The network used in the proposed method builds on Ultra-NeRF~\cite{ultra-nerf}. However, we simplify the physics-enhanced rendering step by employing a more streamlined approach, as detailed in Section~\ref{sec: methods}. Section~\ref{sec: expA} demonstrates the superiority of our technique in preserving image quality, and determines the minimum number of angular views necessary for effective training. Additionally, Section~\ref{sec: expB} evaluates the impact of our rendering approach on ultrasound metrics. Finally, our method achieves a compression ratio of 15:1 for storing model weights compared to the storage required for PW images.

%% file: sections/2_methods.tex
\begin{figure}
    \centering
    \includegraphics[width = \columnwidth]{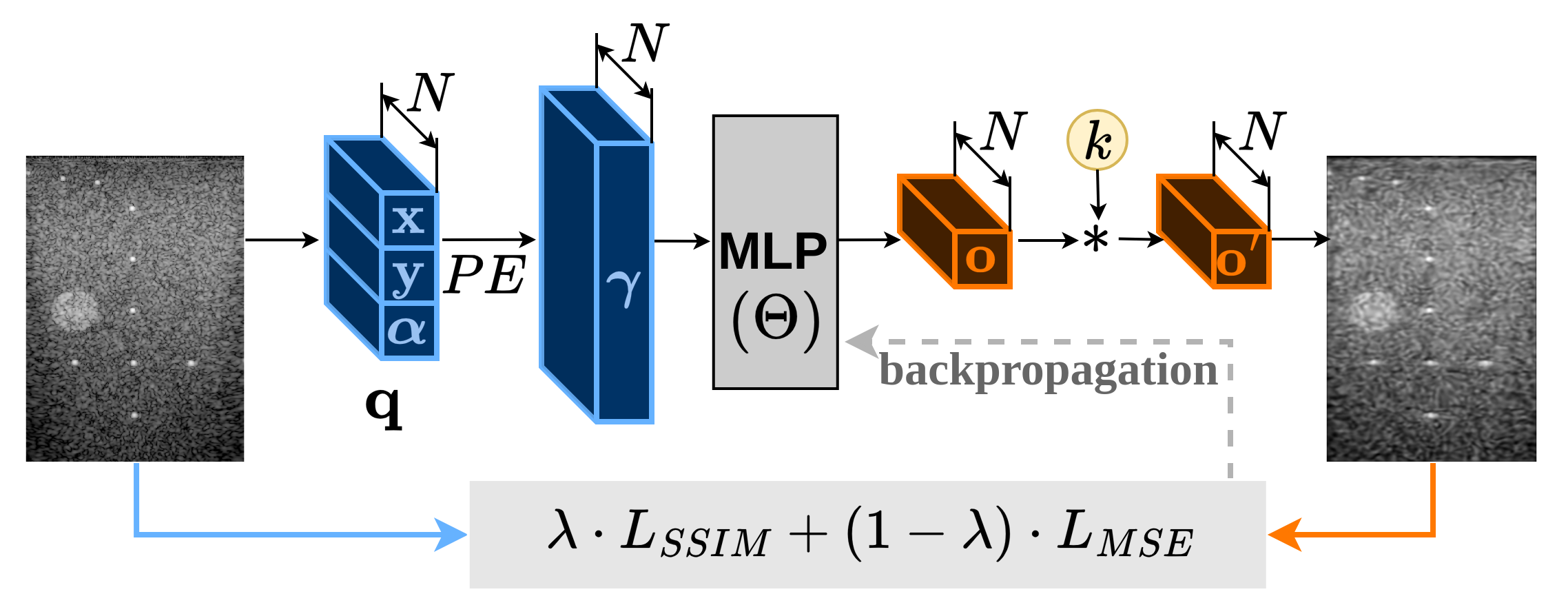}
    \caption{PW angular dependent Implicit Neural Representation.}
    \label{fig: overview}
\end{figure}

The proposed method leverages a Multi-Layer Perceptron (MLP) to represent PW images from multiple angles, see Fig.~\ref{fig: overview} for an overview. By training the MLP on a set of PW images captured from different angles, the network learns to map the coordinates and PW angle given as inputs, to the pixel value that should be observed at that location and from that angle. This process allows the network to infer a continuous function from the discrete image data, effectively enabling smooth interpolation and representation of PW images across various angles.


Let $\qv = \left[ \xv, \yv, \boldsymbol{\alpha} \right]\T$
represent a batch of positional and angular information with dimensions \(3 \times N\), serving as the input to the network. Here, \(N\) denotes the batch size, \(\xv \in \mathbb{R}^{N}\) and \(\yv \in \mathbb{R}^{N}\) correspond to the lateral and axial coordinates, respectively, and \(\boldsymbol{\alpha} \in \mathbb{R}^{N}\) represents the PW angles.

To enhance the MLP's ability to capture high-frequency features in ultrasound data, we employ a Positional Encoding (PE) to transform the input data $\qv$ into higher-dimensional embeddings. Following the approach outlined in \cite{nerf, ultra-nerf}, the embedded MLP's input is formulated as:
\begin{align}
    \boldsymbol{\gamma} = & \left[\qv\T, \sin \left(2^0 \pi \qv\right)\T, \cos \left(2^0 \pi \qv\right)\T, \ldots, \right. \nonumber \\
               & \left. \sin \left(2^{L-1} \pi \qv\right)\T, \cos \left(2^{L-1} \pi \qv\right)\T \right] \T,
    \label{eq:PE_layer}
\end{align}
where $L$ denotes the embedding size and $\boldsymbol{\gamma} \in \mathbb{R}^{(6L+3) \times N}$.

Let $F_{\Theta}$ denote the MLP, where $\Theta$ represents the network parameters. The inference performed by the neural network is expressed as:
\begin{equation}
    F_{\Theta} (\boldsymbol{\gamma}) = \ov,
    \label{eq:intermediate_intensity}
\end{equation}
where 
$\ov \in \mathbb{R}^{N}$ refers to the pixel intensities.

To enhance network training, the proposed method incorporates some light ultrasound data acquisition physics. Specifically, the point-spread function of the probe is modeled using a 2-D blurring kernel. The output of the MLP is convolved with this kernel to generate the final intensity prediction, which is then used to compute the training loss. This convolution process is described by the following equation:
\begin{equation}
    \ov \ast k = \ov',
    \label{eq:final_intensity}
\end{equation}
where $k$ denotes the blurring operator and $\ov' \in \mathbb{R}^{N}$ represents the final predicted intensities.

The loss function, which measures the discrepancy between predicted intensities ($\ov'$) and actual intensities (\textit{Ground Truth (GT)}), integrates both Structural SIMilarity (SSIM) and Mean Squared Error (MSE) metrics, and is defined as:
\begin{equation}
    \lambda \cdot L_{SSIM}(\ov', GT) + (1-\lambda) \cdot L_{MSE}(\ov', GT)
\end{equation}
where $\lambda$ is empirically set to 0.75.

Upon training with multiple PW images from the same field of view but captured at various angles, the network learns to model a continuous function that translates positional and angular information into pixel values. Consequently, it accurately represents the training images and can predict new views based on specified coordinates and angles.

%% file: sections/3_experiments.tex
\subsection{Experimental setup}
In all experiments, the embedding size $L$ for positional encoding is set to 10, and the Multi-Layer Perceptron (MLP) consists of 8 layers, each with 256 neurons. A skip connection is implemented from the embedded input to the fifth layer, and the activation function used throughout the network is ReLU. The total number of trainable parameters is 102,000. This architecture is inspired by NeRF \cite{nerf} and Ultra-NeRF \cite{ultra-nerf}, though it differs from these models in its rendering approach.

The rendering approach employed in this study, as expressed in \eqref{eq:final_intensity}, models the anisotropic point-spread function of the probe. In practice, an anisotropic 2-D Gaussian kernel is used for this purpose. This kernel is derived from the outer product of two 1-D Gaussian kernels, each with a mean of zero. The axial 1-D kernel has a standard deviation of 2 pixels, while the lateral 1-D kernel has a standard deviation of 4 pixels, reflecting the typical lower lateral resolution in ultrasound imaging compared to axial resolution. Consequently, the resulting 2-D kernel has a size of 11 by 11 pixels.

The experiments were conducted using an NVIDIA RTX 4000 GPU, with a batch size of 40,278 for both training and inference. Under these settings, training the model for 10,000 iterations required approximately 50 minutes of wall-clock time.

The datasets used in the experiments are from PICMUS~\cite{PICMUS}, specifically the \textit{Simulated Resolution (SR)}, \textit{Simulated Contrast (SC)}, \textit{Experimental Resolution (ER)}, and \textit{Experimental Contrast (EC)} sets. Each dataset comprises 75 steered plane waves spanning angles from -16° to 16°. Images reconstructed using DAS from each single PW, presented in decibel units with a dynamic range of [-60, 0], serve as the ground truth (\textit{GT}) for both training and testing. The image resolutions are 685 (axial) by 588 (lateral) for the simulated datasets and 857 by 588 for the experimental datasets.

\subsection{Representation Ability Across Different Training Sizes}
\label{sec: expA}
The speed of ultrasound data acquisition depends on the Pulse-Repetition Frequency (PRF). Reducing the number of required PW angles shortens the acquisition process time. To identify the minimum number of angular views necessary for training an effective INR, we assess the model’s sensitivity to the number of training angles.

The model was trained separately on four PICMUS datasets using 14, 25, 38, or 74 periodically selected PW views. Models trained on the \textit{SR} dataset underwent 30,000 iterations, while those trained on the other datasets underwent 10,000 iterations. For comparison with a state-of-the-art method, an Ultra-NeRF model~\cite{ultra-nerf} was trained using 74 \textit{SR} views.

\begin{figure}
    \centering
    \includegraphics[width = 0.45\columnwidth]{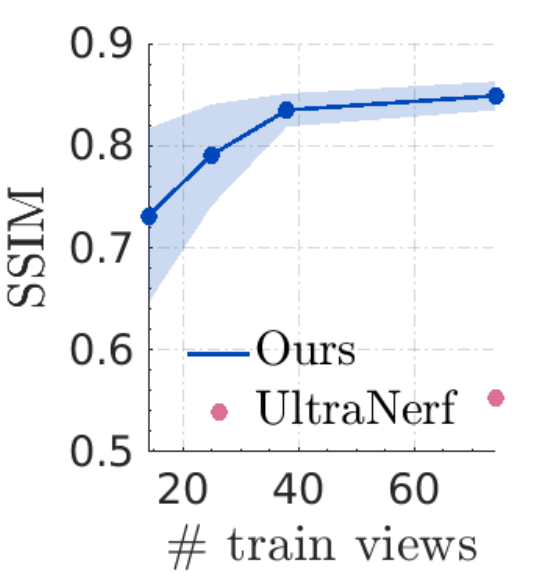}
    \includegraphics[width = 0.45\columnwidth]{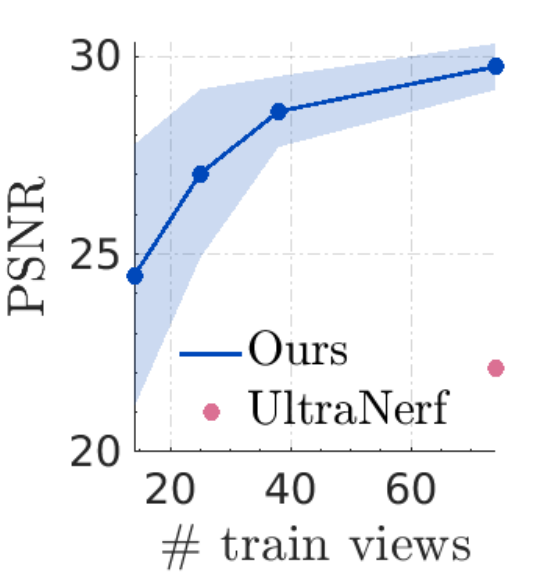}
    \caption{Quantitative comparison of image quality conservation on the \textit{SR} dataset. 30000 iterations, calculated between GT and $\ov'$}
    \label{fig: vs_ultra-nerf}
\end{figure}

\begin{figure}
    \centering
    \includegraphics[width = 0.45\columnwidth]{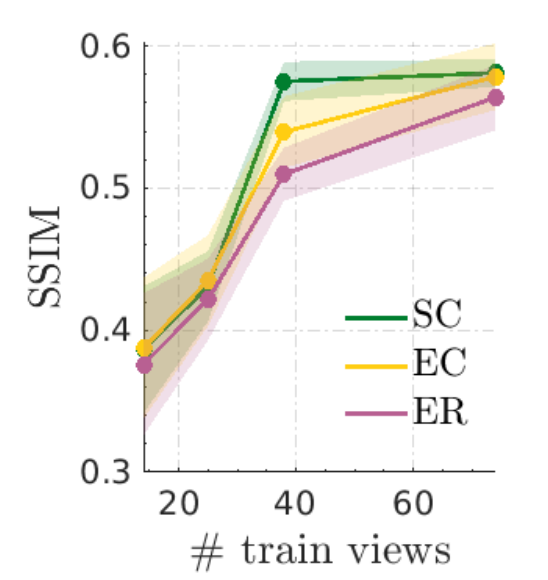}
    \includegraphics[width = 0.45\columnwidth]{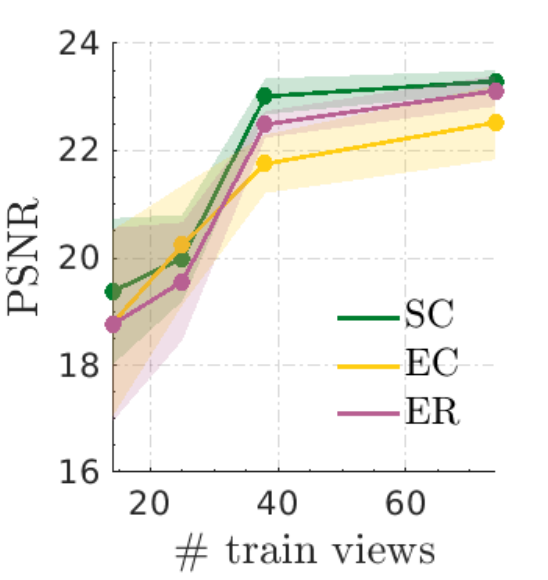}
    \caption{Quantitative comparison of image quality conservation on the \textit{SC}, \textit{ER}, \textit{EC} datasets. 10000 iterations, calculated between GT and $\ov'$}
    \label{fig: CScurve}
\end{figure}

We evaluated image quality using the Structural SIMilarity (SSIM)~\cite{SSIM} and Peak Signal-to-Noise Ratio (PSNR) metrics, comparing the predicted intensities ($\ov'$) with the actual intensities (\textit{GT}) across all 75 angles. The mean and standard deviation of these metrics are shown in Fig.~\ref{fig: vs_ultra-nerf} and Fig.~\ref{fig: CScurve}.

Fig.~\ref{fig: vs_ultra-nerf} presents the results on the \textit{SR} dataset, comparing the performance of Ultra-NeRF~\cite{ultra-nerf} and the proposed method. The comparison indicates that our method surpasses Ultra-NeRF in preserving image quality. Remarkably, even when trained with only 14 views, our method achieves superior SSIM and comparable PSNR to Ultra-NeRF~\cite{ultra-nerf}.

Figures \ref{fig: vs_ultra-nerf} and \ref{fig: CScurve} both illustrate that increasing the number of training views significantly boosts the model's performance. However, the improvement slows down as this number exceeds 38. Therefore, for the subsequent experiments, we have fixed the number of training views at 38.

\subsection{Impact of the Rendering}
\label{sec: expB}
As introduced in Section~\ref{sec: methods} and described by \eqref{eq:final_intensity}, the output of the MLP is rendered to generate the final intensities. This section evaluates both quantitatively and qualitatively the performance of the intermediate predicted intensities ($\ov$) and the final predicted intensities ($\ov'$). The goal is to assess the impact of this rendering approach and determine which type of intensity provides a better representation in terms of conventional ultrasound metrics.

\begin{figure}
   \centering
   \includegraphics[width = \columnwidth]{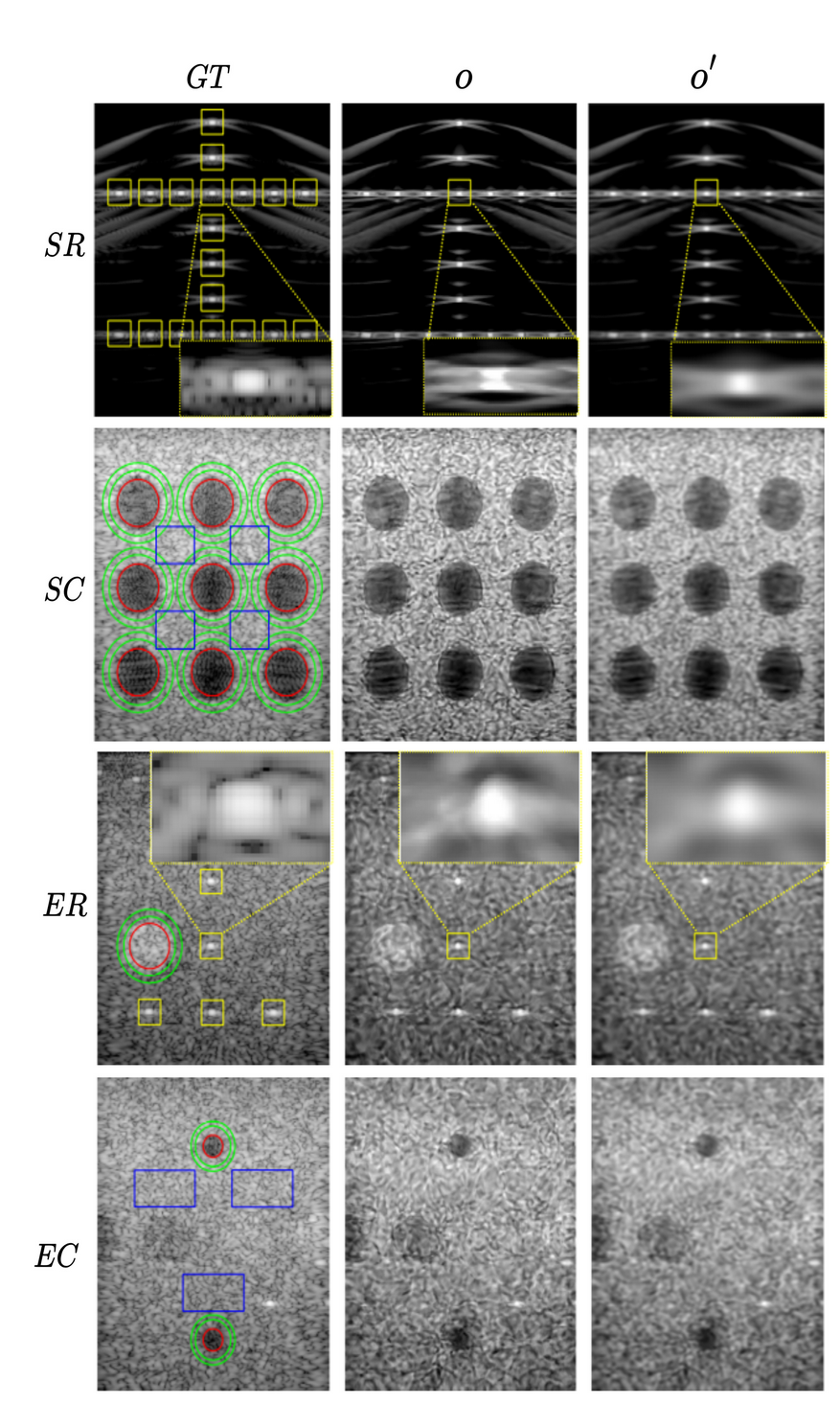}
   \caption{Qualitative results. The model was trained with CS1 (38 views). Regions of interest are outlined in color.}
   \label{fig: imgs}
\end{figure}

For each dataset, the network trained with 38 angular views was employed to predict the orthogonal view. Fig.~\ref{fig: imgs} displays, from left to right, the actual view (\textit{GT}) obtained by DAS from 1 PW, the inferred view ($\ov$), and the rendered view ($\ov'$). For better visualization, a bright scatterer is zoomed in for both the \textit{SR} and \textit{ER} datasets.

Conventional ultrasound metrics are employed to assess the quality of the images. Spatial resolution is evaluated by measuring the -6 dB Full Width at Half Maximum (FWHM) in both the axial and lateral directions on bright scatterers. Contrast is measured using the Contrast-to-Noise Ratio (CNR), calculated as:
$$\mathrm{CNR}=10 \log_{10}\bigg(\frac
{\left|\mu_{\text{in}}-\mu_{\text{out}}\right|^2}{\left(\sigma_{\text {in }}^2+\sigma_{\text {out}}^2\right) / 2}\bigg),
$$
where $\mu_{\text{in}}$ and $\mu_{\text{out}}$ represent the mean pixel values inside and outside the target regions, and $\sigma_{\text{in}}$ and $\sigma_{\text{out}}$ denote the corresponding standard deviations. The CNR is measured in the hyperechoic region for the \textit{ER} dataset and in anechoic regions for the \textit{SC} and \textit{EC} datasets. Additionally, the Signal-to-Noise Ratio (SNR) of the background is calculated as $\mu_\text{ROI} / \sigma_\text{ROI}$, where $\mu_\text{ROI}$ and $\sigma_\text{ROI}$ refer to the mean and standard deviation within the Region Of Interest (ROI). All the measured regions are highlighted in color in the left column of Fig.~\ref{fig: imgs}. Table~\ref{tab: evaluation} presents the average values of each metric for every dataset.

The results depicted in Fig.~\ref{fig: imgs} and summarized in Table~\ref{tab: evaluation} demonstrate the effectiveness of the proposed model in representing PW views and enhancing image quality. Specifically, the intermediate predictions, generated before applying 2-D Gaussian blurring, produce images with sharp edges that improve spatial resolution but at the expense of reduced contrast and lower background SNR. In contrast, the final output ($\ov'$) achieves a superior overall balance by combining the enhanced background SNR and contrast from the rendering process with the improved spatial resolution seen in the intermediate predictions.

\input{tables/quantitative_results}

%% file: tables/quantitative_results.tex

\begin{table}
\centering
\caption{Quantitative comparison. (Best values in bold)}
\begin{tabularx}{\columnwidth}{c|c|>{\centering\arraybackslash}X|>{\centering\arraybackslash}X|>{\centering\arraybackslash}X}
\hline
                             &                              & GT         & $o$           & $o'$           \\ \hline
\multirow{2}{*}{\textit{SR}} & FWHM\_A {[}mm{]}$\downarrow$ &  0.40 & 0.45          & \textbf{0.39}  \\ \cline{2-5} 
                             & FWHM\_L {[}mm{]}$\downarrow$ & 0.81       &  0.71    & \textbf{0.63}  \\ \hline
\multirow{2}{*}{\textit{SC}} & CNR {[}dB{]}$\uparrow$       &  8.08 & 7.77          & \textbf{10.04} \\ \cline{2-5} 
                             & SNR$\uparrow$                & 7.03 & 5.54          & \textbf{8.42}  \\ \hline
\multirow{3}{*}{\textit{ER}} & FWHM\_A {[}mm{]}$\downarrow$ & 0.55       & \textbf{0.47} &  0.50     \\ \cline{2-5} 
                             & FWHM\_L {[}mm{]}$\downarrow$ & 0.89       & \textbf{0.70} & 0.76     \\ \cline{2-5} 
                             & CNR {[}dB{]}$\uparrow$       & 5.63       &  5.72    & \textbf{7.39}  \\ \hline
\multirow{2}{*}{\textit{EC}} & CNR {[}dB{]}$\uparrow$       & 7.74       & 9.17          & \textbf{11.45} \\ \cline{2-5} 
                             & SNR$\uparrow$                & 6.66       &  6.75    & \textbf{8.80}  \\ \hline
\end{tabularx}
\label{tab: evaluation}
\end{table}

%% file: sections/4_conclusion.tex
To the best of our knowledge, this work is the first to employ INRs for representing PW views and for PW angular interpolation.

In terms of storage efficiency, our model's weights, saved in a .npy file, occupy 530 KB, while storing 75 PW images requires 8 MB, resulting in a compression ratio of approximately 15:1. The efficacy of this compact representation is quantitatively assessed using SSIM and PSNR across varying training sizes, and compared against a state-of-the-art method.

The effectiveness of the proposed physics-enhanced rendering is evaluated both qualitatively and quantitatively, through visualization and conventional ultrasound metrics for the task of orthogonal view prediction.

In summary, the proposed lightweight model delivers high-quality PW view representation, and the concise rendering efficiently improves image quality.


%% file: sections/acknowledgment.tex
This work has been supported by the European Regional Development Fund (FEDER), the Pays de la Loire Region on the Connect Talent scheme (MILCOM Project) and Nantes Métropole (Convention 2017-10470).